# Understanding Branly's effect through Induced Tunnelling

Charles Hirlimann

Institut de Physique et Chimie des Matériaux de Strasbourg (IPCMS)

UNISTRA-CNRS (UMR 7504), 23 rue du Lœss BP 43, F-67034 Strasbourg cedex 2, France

**Abstract:** At the end of the nineteenth century Édouard Branly discovered that the electrical resistance of a granular metallic conductor could drop by several orders of magnitude when excited by the electromagnetic field emitted by a distant electrical spark [i]. Despite the fact that this effect was used to detect radio waves in the early days of wireless telegraphy and more recently, studied in the field of granular materials, no satisfactory explanation of the physical origin of the effect has been proposed. In this contribution, we relate the Branly effect to the *induced tunnelling* effect first described by François Bardou and Dominique Boosé [ii].

## 1. *Introduction*

The discovery and characterisation of electromagnetic waves by Hertz at the end of the nineteenth century ignited studies on the interactions of these waves with materials. One of the first major findings was that of the radioconduction of granular conductors by Branly in 1890 [1]. He found that when an insulator tube filled with metal filings is submitted to a DC potential difference and it receives transient electromagnetic waves generated by an electrical spark at some distance away, its electrical resistance drops by several orders of magnitude. Many aspects of this effect were revealed in other work during the following decade. It has been established firmly that the presence of a thin resistive layer between the grains is necessary to achieve high resistance before exposure to the electromagnetic waves. As the effect does not occur with noble metal grains cleaned of any surface contaminant, the nature of the resistive layer was recognised as an oxide or vacuum by Dorn [iii]. It has been observed previously that a permanent and unique conductance percolation path is the reason for the drop in resistance of the granular medium [iv,v], which led to the understanding that the welding of the grains was the result of radioconduction or the coherer effect. Because of this welding of the grains the effect is non-reversible, although the electrical path can be broken easily by gently striking the tube containing the filings, which restores the original large resistance of the granular medium. This effect was used by Lodge [vi,vii] to devise a practical radio detector that ensured the pioneering developments of radiotelegraphy. Pressing the grains against each other reduces the original resistance by reducing the thickness of the oxide layers and thus, increasing the sensitivity of the radioconductor to the external excitation [viii]. It has always been observed that the effect is inherently noisy and that the final resistance fluctuates within a rather large range of results [ix].

Above some critical applied DC potential, the resistance of a granular medium drops without any external excitation [3]. This effect, which must not been confused with the Branly effect, even though in recent times it has been called the continuous Branly effect, is what we wish to discuss here [x]. Ashkinass studied the resistance drop of a unique contact between two tin needles and found a critical voltage of 0.2 V, independent of the current [xi]. Guthe and Trowbridge recorded the U(I) curve of the contact between two metallic spheres and showed that without any external excitation, it exhibits a critical voltage P that characterises the resistance transition of the metal-insulator-metal (MIM) system. They found the value P = 0.23 V for the DC potential drop between two centimetre-sized steel balls. As can be seen, these experiments involve only one barrier of potential made of the oxide of the two metals in contact. Therefore, we understand now that this "continuous Branly effect" is the "classical" tunnelling of an electron through a barrier of potential with an energy level of the order of 0.2 eV, under the experimental situations that have been explored. The resistance of the device drops as soon as the applied voltage accelerates the conducting electrons to an energy that is larger than the barrier height.

Interest in the effect under its various forms resumed during recent decades in relation to a renewed interest in granular media. The uniqueness of the conduction path, which results from a percolation of the welding of the grains through the granular medium, was confirmed using infrared optical imaging [xii]; the welding of the grains has been firmly and quantitatively established. Careful experiments in the CW mode on conduction between metal spheres, has shown through sophisticated modelling that high electric current through contacts with 100-nm diameters produces a temperature as high as 1050 °C, which is large enough to ensure welding between the metallic conductors [10].

The Branly effect itself, in which a direct voltage of less than the critical voltage is applied to the granular medium and an electromagnetic pulse dramatically changes its total resistance, was studied in ordered lattices of lead beads by Dorbolo et al. [xiii].

Briefly, we consider here the radio conduction effect discovered by Branly in which oxidised metal particles set under a DC voltage are welded together when an electromagnetic pulse is applied. Despite the great wealth of results that have been collected over more than a century, no physical understanding has yet been proposed regarding the physical origin of this radioconduction effect. This paper proposes an explanation based on the "induced tunnelling effect" that has been theoretically described recently and evidenced in an optical experiment [2,xiv].

## 2. *Experimental results*
### 2.1 *Radioconduction*

The discussion can be first simplified by noting that the granular structure of the conducting medium and the related percolation effects are not relevant to the radioconduction effect, as Branly himself showed in a very elegant experiment performed with a couple of conductors separated by a thin oxide layer [xv]. We can assert then that the Branly effect is governed by the electrical tunnel effect that occurs in what is nowadays called a MIM structure [16]. In a typical experiment, Branly observed that when a spark in the surroundings emits electromagnetic waves, the electrical resistance of a granular medium drops. In these experiments a Daniell battery applies a U = 1 V DC potential difference to the medium, which comprises iron filings that are compressed slightly in a glass tube; a galvanometer measures the current flowing through the filings. Extreme drops in the resistance were measured at the time an electrical spark was ignited in the vicinity of the experiment. A typical measurement of change of resistance of the tube was from $R_0 = 10^6$ Ω to R = 100 Ω; a change of four orders of magnitude. According to Ohm's law, the original current flowing through the unperturbed granular medium in that case is $i_0 = 10^{-6}$ C.s$^{-1}$, which corresponds to a flow of electrons of $n_0 = 6.10^{12}$ s$^{-1}$. This current increases up to $i_f = 10^{-2}$ C.s$^{-1}$ = 10 mA when the Branly effect occurs and the granular medium starts conducting. In this conducting state, the flow of electrons becomes $n_f = 6.10^{16}$ s$^{-1}$.

### 2.2 *The CW Branly effect*

Recent experiments [10] have been performed very carefully with chains of stainless steel balls in the CW regime with no external excitation. In these experiments, the steel balls are separated by nanometre-thick oxide layers acting as barriers of potential for the conducting electrons. By applying a DC voltage to the set of contacting balls, the authors found a value of P = 0.4 V per contact (equivalent to per barrier of potential) for the critical voltage above which the resistance of the chains of balls drops. Below the critical value, the kinetic energy of the electrons in the balls is less than the height of the barrier and thus, conduction is governed by the tunnelling of electrons through the barriers and the electrical resistance is high. Above the critical voltage, the kinetic energy of the conducting electrons is greater than the barrier height and thus, they behave like free electrons in a metal; the conductivity increases by orders of magnitude and the resistance of the device drops. The value of the critical voltage is in very good agreement with the historical observations referred to above. These authors were able to establish through modelling that the drop of resistance is a consequence of local heating with a peak temperature of 1050 °C; high enough to allow for the local destruction of the barrier of potential and for the local welding of the steel balls. Therefore, when the current flowing through the MIM contacts of a coherer is forced to increase, a threshold can be reached such that a localised welding of the metallic parts occurs and hence, the resistance of the contacts drops irreversibly.

At this point, we understand that the physics at the core of the Branly radioconduction effect, which was used previously for detecting electromagnetic pulses, is based on the transmission of electrons through a barrier of potential equal to a few hundred meVs. These barriers are fabricated from nanometre-thick layers of oxide separating two non-noble metal parts (e.g., filings, spheres, wires, tips). We also understand that the potential difference applied to the tube of filings in Branly's experiments was always such that the voltage applied to one contact was always less than the critical voltage above which the electrons would have energy greater than the barrier height. The mode of electrical conduction in Branly's experiment was therefore classical tunnelling, leading to high electrical resistances of the medium and low electron flows.

## 3. *Discussion*

### 3.1 *Induced tunnelling*

Once the radioconduction effect is understood, the question that remains to be addressed is why the electrons in the metal parts in the filings tube behave as if they had an energy larger than the height of the barrier of potential that separates the grains, at the time when a spark emits an electromagnetic pulse in the vicinity of the experiment.

Let us first consider the order of magnitude of the energy of the electromagnetic field emitted by the exciting spark that could be deposited on the electrons in the metal particles. Branly used the classical setup designed by Hertz for generating electromagnetic pulses [1]. The primary source of energy was a chemical battery. A Ruhmkorff induction coil was used in the experiment to generate trains of sparks emitting gigahertz electromagnetic (em) waves. The energy in the spark plasma is estimated to be of the order of 10 mJ [17] and because of the poor yield of the conversion of this energy into em waves, these waves could not be detected at distances greater than a few metres.. In Branly's experiment, the spark generator was placed at a distance of 20 m from the experiment. At this point, two regimes must be distinguished depending on whether the spark generator is close to the experiment. In the experiment by Dorbolo et al. [13], for instance, the distance does not exceed one metre; in that situation, the energy transferred by the em wave to the electrons in the metal parts is sufficient for them to energetically overcome the barrier energy height. This does not differ from the continuous Branly effect described above, in which high temperatures are reached and welding of the contacts occurs when the flow of electrons is sufficiently large.

This energy transfer does not stand in experimental situations where the distance is larger and where the energy deposited by the em wave bursts becomes much smaller than the barrier height. It is the purpose of this paper to propose the "induced tunnelling effect" as the physical mechanism responsible for induced conduction when the energy of the electrons in the metal parts is smaller than the barrier of the potential that separates them.

In a work published in 2001, Bardou and Boosé established theoretically that the probability of a particle tunnelling through a barrier of potential could be increased significantly by gently striking the particle at the time when the centroid of its wave packet is reflecting on the barrier [2]. The one-dimensional model calculation they performed stands for a square potential barrier of height $U_0$ and an electron with energy E, which is described as a wave packet (Fig. 1a). It was shown that a small momentum transfer to the particle $\Delta p_x$ (such that the energy transfer $\hbar^2 \Delta p_x / 2m \ll E$, m being the mass of the particle) at the time of its reflection on the barrier of potential has the dramatic effect of projecting its wave functions on the full set of its Eigen-functions. This full set of wave functions comprises the stationary wave functions of the undisturbed Schrödinger equation. Therefore, the result of the momentum transfer is such that the weight of the continuum of wave functions, corresponding to the particle energies lying in the continuum far above the height of the barrier of potential, becomes dominant in the final wave function. As the probability of the transmission of the particle through the barrier is obtained by integrating the contributions of all the wave functions, the effect of the momentum transfer is to increase the transmission probability because of the large contribution of the passing wave functions (Fig. 1b). These authors have shown that the excitation of the particle at the time of its reflection on a barrier of potential increases it transmission probability by several orders of magnitude. Specifically, below the transfer limit, the transmission T of the barrier is proportional to the square of the momentum transfer $T(\Delta p_x) \Box (\Delta p_x)^2$. As a consequence, the induced tunnelling effect does not depend on the sign of the momentum transfer, i.e., the effect is the same whether the particle is pushed towards the barrier or pulled away from it. This is significantly different from the regular tunnel effect in which the transmission probability varies as $T(E) \propto \exp(-a\sqrt{U_0 - E})$, in which $a$ is a constant of proportionality that depends on the thickness of the barrier and hence, it increases when the particle is pushed towards the barrier (the energy increases) and decreases when the particle is pulled away (the energy decreases).

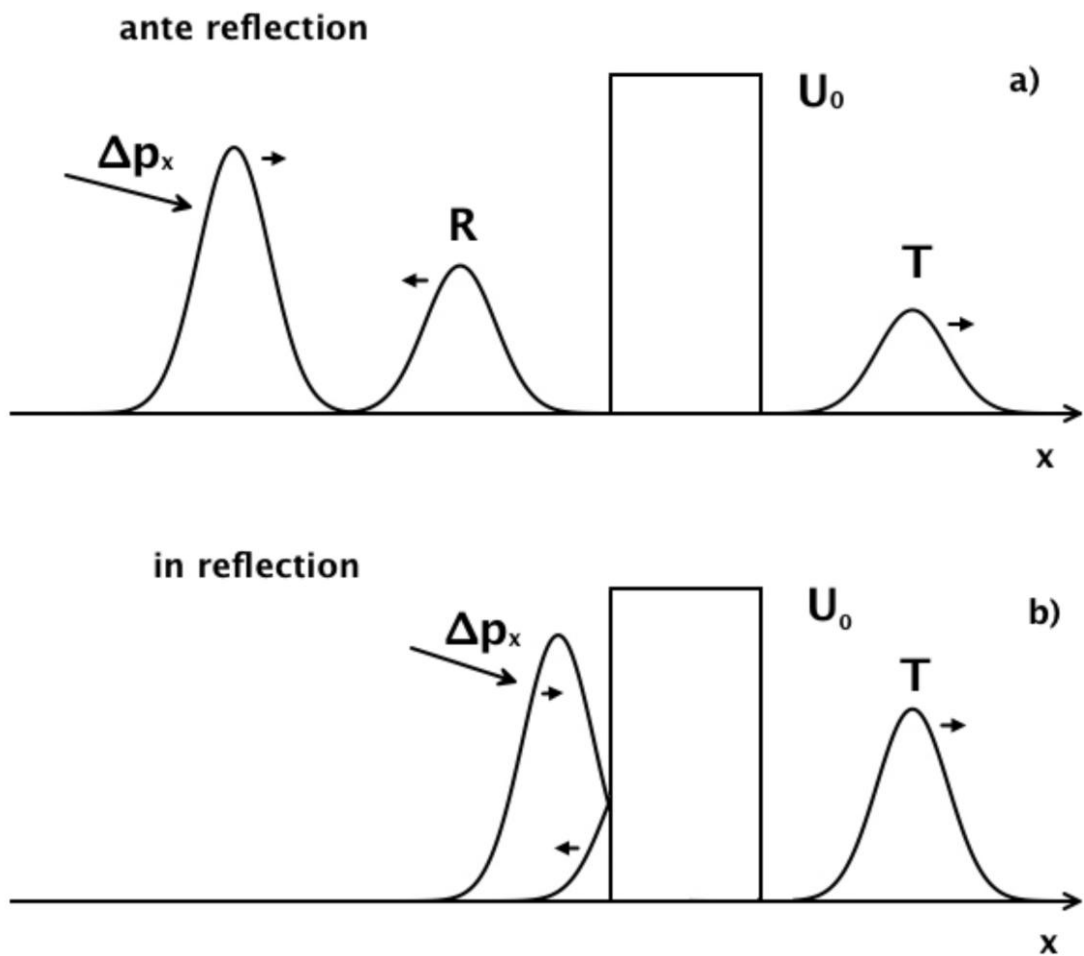

**Figure 1** Sketch of the induced tunnelling effect. a) Classical one-dimensional quantum tunnelling. One particle,

shown as a wave packet on the left, hits the barrier of potential of height $U_0$. The particle has a probability $T$ given in the main text, to be transmitted through the barrier and a probability $R$ to be reflected. If a small momentum transfer $\Delta p_x$ is transferred to the particle before it reaches the barrier, its energy changes according to the sign of the momentum transfer $\hbar^2 \Delta p_x / 2m \ll E$ and the probabilities change only slightly. b) Induced tunnelling. If the momentum transfer occurs at the time the incoming and the reflected wave functions are superimposed, then the transmitted probability is greatly enhanced. This enhancement is quadratic with the momentum transfer and therefore, it does not depend on the sign of the kicking of the particle. This enhanced probability is due to the projection of the particle wave function on its full set of Eigen-functions and as a consequence, it is due to the large contribution of the passing wave functions.

One important feature of the induced tunnelling effect lies in the fact that it reaches its maximum amplitude at a time that corresponds to the exact reflection on the barrier, when one half of the incoming wave packet interferes with the reflected other half. It is therefore important to estimate the duration of the reflection of a particle on the barrier of potential, because the duration of the momentum transfer has to be of the same order of magnitude or less. This transient interference process is, of course, not instantaneous and its duration is related to the time it takes for an evanescent wave to establish inside the barrier of potential [18]. This reflection delay is quite general in wave physics and in optics, it is known as the Goos-Hänchen effect when total reflection occurs [19]. A textbook calculation shows that the duration $\tau$ of the reflection of a particle on a barrier of potential can be written as:

$$\tau = \frac{2m}{\hbar k \sqrt{\kappa_0^2 - k^2}}, \quad \kappa_0^2 = \frac{2mU_0}{\hbar^2}, \quad k^2 = \frac{2mE}{\hbar^2},$$

$$\tau = \frac{0.66}{\sqrt{E(U_0 - E)}} \quad \text{in fs when m is the mass of an electron.}$$

(1)

Under physical conditions where the height of the barrier is chosen to be $U_0 = 0.5$ eV, representative of the critical potentials found in the literature, the energy of the particle is chosen here to be $E = 0.25$ eV and this time delay is equal to 2.65 fs. To make the point more precise, Fig. 2 shows the variation of this delay in the reflection on the barrier of potential of an electron as a function of the energy of this electron.

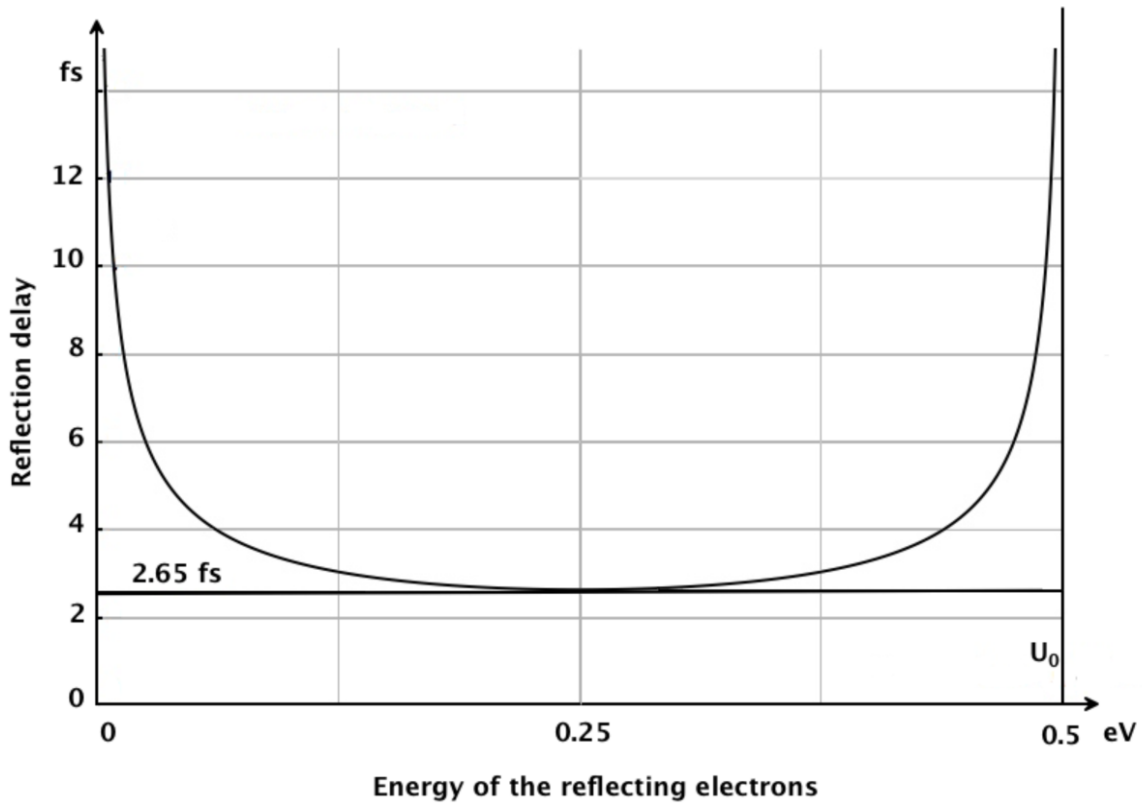

**Figure 2** Variation of the reflection delay for a particle reflecting on a barrier of potential with height $U_0$ as a function of its energy. As can be observed, the reflection delay increases greatly for particles with either energy close to the barrier height or for slow particles with low energy, which is less intuitive. These relatively large durations are favourable to the perturbation of a reflecting particle by an incoming short-lived excitation.

The very interesting point that arises from the calculation is that both high and low energy electrons have rapidly increasing sticking delays when their energy increases or decreases, respectively. This situation is highly favourable to their efficient kicking through an external excitation. Outside of this sticking time interval, any momentum transfer is inefficient and equivalent to a very small energy change of the impinging electron in the regular tunnelling effect [2]. Therefore, the external excitation of a particle reflecting on a barrier must contain dynamical structures on the time scale as the sticking delay. We shall now scrutinise the temporal structure of the electromagnetic pulse that is emitted by a spark and verify whether it could act as the external excitation.

### 3.2. *Electrical sparks' temporal structure*

As early as at the beginning of last century, Abraham and Lemoine discovered experimentally that the duration of an electric spark is less than one nanosecond [20]. Recently, Descœudres made exhaustive spectral measurements of electrical sparks and found a bandwidth of the order of 400 nm in the visible range [21]. An electrical spark is the result of electronic de-excitation of electrons and ions in gas plasma and the various emitted electromagnetic waves do not have any reproducible phase relationship, i.e., the emission is incoherent in nature. As with any incoherent emission, spikes or bursts of emission intensity occur erratically. Because from the measured spectral width of a spark emission, a coherence time as short as one femtosecond can be inferred from the standard Fourier time-frequency relationship [22], it seems clear that the condition of shortness in the momentum transfer to the electron can be fulfilled by the electromagnetic field emitted by an electric spark. It is therefore assessed that induced tunnelling can result from the kicking of the electrons in a metal particle reflecting onto a barrier of potential.

At this point, it becomes clear that the induced tunnelling effect gives the appropriate framework for an interpretation of the Branly effect. In a granular oxidised metallic medium, microscopic grains are isolated electrically from one another by a nanometre-thick metal oxide layer, forming a large population of MIM's. When a voltage is applied to the medium, electrons are accelerated and they reflect on the barriers of potential that separate the grains of the medium. At the time of the reflection, these electrons can be kicked forwards or backwards by short electromagnetic bursts present in the external electromagnetic field emitted by a spark, as a small momentum is transferred to them. The large transmission enhancement induced by the momentum transfer produces an increased electrical current, which for some events becomes sufficiently large to permit local heating in the metal grains and thanks to the Joule effect, leads to the welding of pairs of grains. When a percolation path has formed due to the welding of the grains, the electrical resistance of the medium drops changing from an exponential dependence on the applied voltage to a linear one (Ohm's law).

This theory should be easy to falsify. For instance, excitation of a coherer by a smooth, long-lasting electromagnetic field should be inefficient because a fast momentum transfer is needed, one that lasts no longer than the reflection time of the electrons on the oxide barriers in the framework of the induced tunnelling effect. Gold dots deposited on a resistive surface and separated by 1-nm-thick vacuum (air) barriers should behave as instantaneous coherers, because the resistance drop would be controlled only by the induced tunnelling effect and no welding between the dots would be possible in that experimental situation.

## 4. *Conclusion and perspective*

The drop of electrical resistance of a granular oxidised medium under the influence of an external incoherent electromagnetic burst has not received a satisfactory physical explanation for more than a century, despite the fact that it was used by Lodge [7] to design the coherer, the first sensitive electromagnetic wave detector and despite it making possible the first long-distance wireless telecommunication by Marconi in 1895 [23]. In this paper, we have discussed the possibility that the induced tunnelling effect, which was discovered theoretically by Bardou and Boosé and experimentally verified by Hirlimann et al., could be responsible for the Branly effect, which was at the heart of early wireless telecommunications.

This physical interpretation given to the Branly effect can be applied to various other manifestations of resistance changes under external excitation of a granular medium that have been observed. In 1898,

Auerbach demonstrated that a coherer could be made conducting by an acoustic excitation in the audible range of the spectrum [24]. As mentioned in [2] and described more precisely in [14], there exist other ways to produce induced tunnelling, which rest on the fact that particles receive additional momentum when they impinge on a moving potential barrier [25]. Therefore, acoustic waves, by inducing vibration in the tunnel barriers between the grains of a coherer, could be responsible for an induced tunnelling. Moreover, acoustically induced tunnelling could also be at the very heart of the microphone effect between carbon grains, which has been abandoned by the telephone industry only during the last decade [26]. Related to these experiments is the increase of resistance observed when metallic filings are replaced by what is nowadays called semiconductor grains. A typical experiment was performed by Ashkinass using $PbO_2$; when excited externally by a strong electrical spark, the resistance of a tube filled with $PbO_2$ powder increases by several orders of magnitude [11]. Here again, tunnelling induced by an external electromagnetic wave burst could be responsible for a charge (electrons and holes) migration on both sides of the tunnels separating the grains. The charge separation, by creating a region of space, would increase the thickness of the tunnels and therefore, increase the resistance of the medium.

The new concept of induced tunnelling has a strong unifying power on the various effects that occur in situations where quantum tunnels are present. As has been described above, the effect is sensitive to any excitation from surrounding mechanical and electromagnetic noise. The result is a noisy intrinsic behaviour of any device that can exhibit the Branly effect. As a matter of fact, coherers used in the early stages of radio telegraphy have been difficult to handle because of the large variation of their output; carbon granular microphones are known to exhibit specific "carbon hiss" noise and large noise levels are observed in the MIM structures [11, 27] used nowadays in memory devices. Since the first measurement of the electrical resistance of a single $C_{60}$ molecule in 1995 [28], the growth in the number of experiments in the field of molecular electronics has been exponential. A large number of these experiments rely on the use of tunnel effects and should therefore be sensitive to induced tunnelling. Thus, care should be taken to strictly isolate the experimental setups from external excitations, either electromagnetic or acoustic. Last but not least, it has been known for a long time that alpha radioactivity [29,30] is due to the tunnelling of a hydrogen nucleus out of large and unstable nuclei. This consideration opens the opportunity of exploring the possibility for gamma rays with energy less than the barrier height to induce tunnelling of hydrogen particles; hence, decreasing the radioactive lifetime of the emitting nuclei. This could also be explored in case of beta emission.

**In memoriam.** This work is dedicated to François Bardou†.
**Acknowledgement**. The author thanks Romeo Michelangelo for reproducing Branly's experiments in the course of this work. Dr Jean-François Morhange performed a very careful reading of the manuscript, many thanks to him.